\documentclass[onecolumn,amsmath,amssymb,floatfix]{revtex4}
\usepackage{graphicx}
\usepackage{dcolumn}
\usepackage{bm}
\usepackage{hyperref}
\usepackage{color}
\usepackage{epsfig}
\usepackage{subfigure}  
\usepackage[utf8x]{inputenc}

\newcommand{\beq}{\begin{equation}}
\newcommand{\eeq}{\end{equation}}
\newcommand{\begg}{\begin{gather}}
\newcommand{\eegg}{\end{gather}}

\begin{document}

\title{Self-force driven motion in curved spacetime}

\author{
Alessandro D.A.M. Spallicci$^1$\footnote{spallicci@cnrs-orleans.fr~http://lpc2e.cnrs-orleans.fr/$\sim$spallicci/},
Patxi Ritter$^{1,2}$\footnote{patxi.ritter@cnrs-orleans.fr~http://lpc2e.cnrs-orleans.fr/$\sim$pritter/},
Sofiane Aoudia$^3$\footnote{aoudia@aei.mpg.de~http://www.researchgate.net/profile/Sofiane\_Aoudia}}
\address
{
$^1$Universit\'e d'Orl\'eans\\ 
Observatoire des Sciences de l'Univers en r\'egion Centre (OSUC) UMS 3116 \\
Centre Nationale de la Recherche Scientifique (CNRS)\\
Laboratoire de Physique et Chimie de l'Environnement et de l'Espace (LPC2E) UMR 7328\\
3A Avenue de la Recherche Scientifique, 45071 Orl\'eans, France\\
$^2$Universit\'e d'Orl\'eans\\ 
Laboratoire de Math\'ematiques - Analyse, Probabilit\'es, Mod\`elisation - Orl\'eans (MAPMO) UMR 7349 \\
Rue de Chartres, 45067 Orl\'eans, France\\
$^3$Max Planck Institut f\"ur Gravitationphysik, A. Einstein\\
Am M\"uhlenberg 1, 14476 Potsdam, Deutschland
}

\date{16 May 2014}

\begin{abstract}
We adopt the Dirac-Detweiler-Whiting radiative and regular effective field in curved spacetime. Thereby, we derive straightforwardly the first order perturbative correction to the geodesic of the background in a covariant form, for the extreme mass ratio two-body problem. The correction contains the self-force contribution and a background metric dependent term. 
\end{abstract}

\maketitle
{\small
 }

\section{The two-body problem and the perturbation method}

For the two-body problem in general relativity, it is widely known that there aren't exact solutions. For studying a real case, we would need 
to consider two masses embedded in a single field (metric), but there isn't any available tool to perform such a desirable computation. At the easy end of the difficulty scale, analytic solutions are available when the mass of the smaller body - if a smaller body can be identified - is neglected. In this case, general relativity dictates the physics via the geodesic equation. 
Between the real and the oversimplified ends, several approaches lie. One of these approaches, Numerical Relativity (NR), although having  scored impressive results, is still partially limited by technical difficulties (initial conditions, mass ratio of the two bodies, duration of the orbital evolution, computing power requirements, etc.). Further, it is not best suited to physical understanding. 
 
Leaving behind us the partially fledged NR, we are forced to enter into the world of approximation techniques dealing with the simplified (linearised) version of the Einstein equation. 
The approximation methods examine different features of the two-body problem at the price of narrowing down their applicability to this or that domain. 
For instance, in strong field and high velocity regime, post-Newtonian (pN) methods cease to be highly reliable. Based on the pN framework, an improvement is offered by the Effective One Body (EOB) approach. EOB reduces the matter at hand to a single body moving in an effective potential. To which extent the EOB improvement goes, it is subject of exploration. Anyway, the EOB is not a self-standing approach, as it is calibrated through the parameters coming from NR, pN and lately perturbation methods. With the latter, a hard, but now manageable, challenge considers both masses for any field and velocity, if one of the two masses is much smaller - but still existing - than the other, and it shrinks to a point. This third approach will be pursued herein. 
 
In this scenario, the small body reacts to its own field-mass and to the radiation emitted. 
Back-reaction to the own field-mass for the small body is to be interpreted solely as induced  by the presence of another body. In other words, a single body - infinitely - far away from any other gravitating body or external influence of any sort, will not experience any interaction with its own field and mass. Self-force is the back-action of a body to its mass, motion and radiation via the intermediate role of an external field.     
      
For the smaller mass, the shrinkage to a point is not painless. The concept of gravitating point-mass is foreign to full general relativity, as its expression through a distributional stress-energy tensor on the world-line leads to inherent contradictions. In linearised relativity, a gravitating point-particle is acceptable, but divergences will be associated to the particle. After all, the difficulties arising from the infinitesimal size are to be traded against the simplifications obtained by neglecting the internal structure.          

The problem of motion for point particles has been tackled by concurring approaches all yielding the same result.
The solution for a massive point-particle moving in a strong field and for any velocity was indeed derived in 1997 by Mino, Sasaki and Tanaka \cite{misata97}, Quinn and Wald \cite{quwa97}, in the de Donder \cite{dd21} (harmonic) gauge\footnote {The harmonic gauge was also proposed by Lanczos \cite{la22}}, around an expansion of the mass ratio, $m/M$, $m$ being the small mass and $M$ the large one. 
The main result has been the identification of the regular and singular perturbation components and their playing and not-playing role to motion, respectively. The resulting equation has been baptised MiSaTaQuWa from the first two initials of its discoverers. A practical recipe, based on spherical harmonics and dealing with  the divergences coming from the infinitesimal size of the particle, was later conceived by Barack and Ori \cite{baor00}. 

The MiSaTaQuWa approach may be intuitively viewed. The particle crosses the curved spacetime and thus generates gravitational waves. These waves are partly radiated to infinity (the instantaneous or direct part) and partly scattered back by the black hole potential (the tail part), thus forming tails which impinge on the particle and give origin to the self-force. 
Alternatively, the same phenomenon is described by an interaction particle-black hole generating a field which behaves as outgoing radiation in the wave-zone and thereby extracts energy from the particle. In the near-zone, the field acts on the particle and determines he self-force which impedes the particle to move on the geodesic of the background metric.
From these works, it emerges the splitting between the instantaneous and tail components of the perturbations, the latter  acting on the motion. Unfortunately the tail component can't be computed directly, but as a difference between the total and the instantaneous components. Detweiler and Whiting \cite{dewh03} have shown an alternative approach, not any longer based on the computation of tails, but stemmed from a geodesic vision of the motion. We shall make use of the DeWh approach to identify the radiative part of the perturbations.         

A comprehensive introduction to mass and motion in general relativity has appeared \cite{blspwh11}. This paper echoes the same intendment, focusing on the self-force for point particles in a gravitational field and on how the geodesic deviation arises in perturbed spacetime. The question we pose is: what is the difference between the motion of a particle in an unperturbed background metric and the same particle being affected by its own field-mass and the radiation emitted? 
We provide two answers to the question: one in the main body of the paper, the other in the appendix.  

The topic attracts growing interest beyond the Capra Meeting community\footnote{http://www.cnrs-orleans.fr/osuc/conf/}, since impacting on the successful accomplishment of an SLI (Space Laser Interferometry) mission like LISA\footnote{https://www.elisascience.org/} for the detection of gravitational waves emitted by EMRI (Extreme Mass Ratio Inspiral) sources.
We use the signature convention (- + + +).

\section{The radiative part of the perturbations}

On the footsteps of Dirac's work \cite{di38}, Detweiler and Whiting \cite{dewh03, po11, popove11} have proposed a novel approach to the self-force. The singular term (from the perturbation field) 
$Sing $, that is the mean of the advanced and retarded terms, is time-reversal invariant, {\it id est} incoming and outgoing energy are equal.  It is known that in flat spacetime, the radiative term is obtained by subtracting the singular from the retarded term. The latter is singular, non time-reversal invariant, and shows that the system is losing energy by radiating outward. The subtraction cancels out the singularity at the particle, without any other consequence. Indeed, the singularity is isotropic and it does not exert any force on the particle. It remains only the radiative term to act upon the particle given by

\[
Rad = Ret - Sing = Ret - \frac{1}{2}[Ret + Adv] = \frac{1}{2}[Ret - Adv]~~.
\]      

\begin{figure}
\centering
\begin{minipage}[c]{0.45\textwidth}
  \includegraphics[width=6cm]{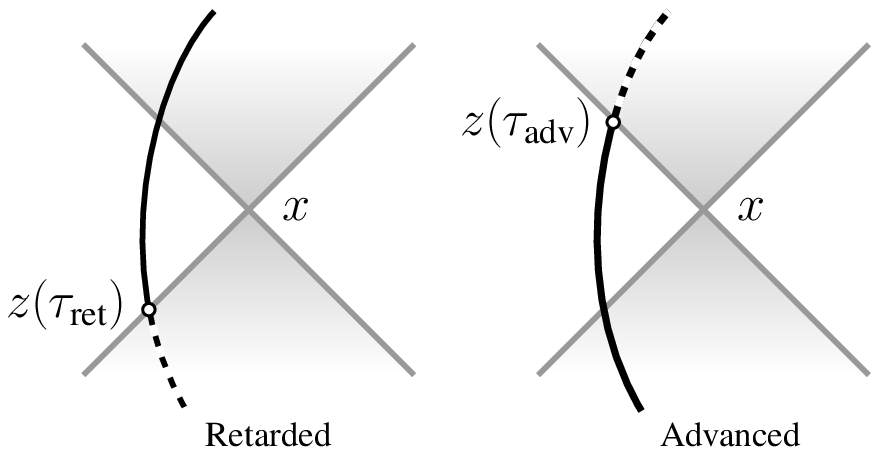}
  \caption{Retarded and advanced terms (dotted line). }
  \label{fig1}
\end{minipage}%
\quad
\begin{minipage}[c]{0.45\textwidth}
  \includegraphics[width=2.6cm]{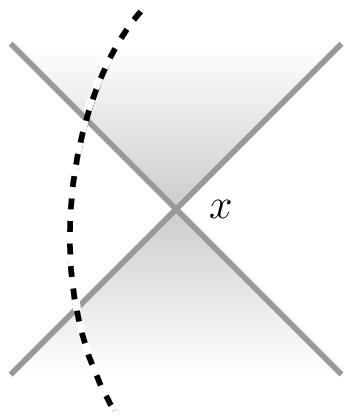}
  \caption{The H term (dotted line). }
  \label{fig2}
\end{minipage}%

\end{figure}

\begin{figure}
  \centering
\begin{minipage}[c]{0.45\textwidth}
  \includegraphics[width=3cm]{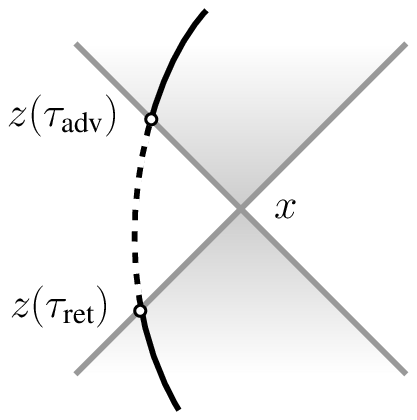}
  \caption{Singular term (dotted line).}
  \label{fig3}
\end{minipage}%
\quad
\begin{minipage}[c]{0.45\textwidth}
    \includegraphics[width=3cm]{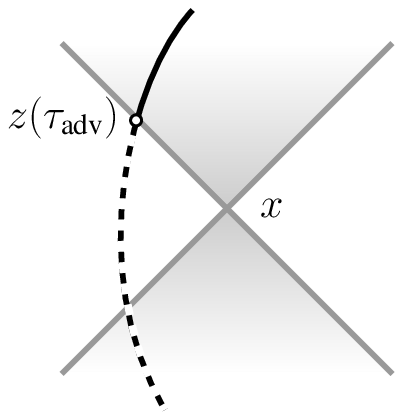}
  \caption{Radiative term (dotted line).}
  \label{fig4}
\end{minipage}

\end{figure}

In curved spacetime, Figs. \ref{fig1}-\ref{fig4}, at a given point $x$ the retarded term depends upon the particle's history before the
retarded time $\tau_{\rm ret}$; the advanced term depends upon the particle's history after the advanced time $\tau_{\rm adv}$.
The singular term depends then upon the particle's history during the
interval $\tau_{\rm ret}<\tau<\tau_{\rm adv}$. 
The straight transposition of the subtraction $Ret - Sing$ to curved space determines still a singularity-free quantity, but the latter  depends upon the contributions from inside of the light cone, past and future. 
The dependence on the future is patently non-causal. The circumvention of this riddle passes through the inclusion of an additional, purposely built, function $H$   

\[ 
Rad = Ret - Sing = Ret - \frac{1}{2}[Ret + Adv - H] = \frac{1}{2}[Ret - Adv + H]~~,
\]      
where the {\it ad hoc} function $H$ is defined to agree with the advanced term when the particle position is in the future of the evaluation point, thereby cancelling the $Sing$ term   (the $Ret$ term is zero, for $\tau>\tau_{\rm adv}$). Finally, we have 

\[
Rad_{~\tau>\tau_{\rm adv}} = 0~~.
\]
 
Instead, $H$ is defined to agree with the retarded term when the particle position is in the past of the evaluation point, also cancelling the $Sing$ term  (the $Adv$ term is zero, for $\tau<\tau_{\rm ret}$). Finally, we have 
\[
Rad_{~\tau <\tau_{\rm ret}} = Ret ~~.
\] 

Further, $H$ differs from zero at the intermediate values of the world-line outside the light-cone, between $\tau_{\rm ret}$ and 
$\tau_{\rm adv}$. Thus, the radiative component includes the state of motion at all times prior to the advanced time, and it is not a representation of the physical field but rather of an effective field. Nevertheless, $H$ goes to zero when the evaluation point coincides with the particle position. Figures (\ref{fig1}-\ref{fig4}) show the various terms ($\tau$ is the proper time, $x$ the evaluation point, and $z$ the particle position). 

The perturbation $h_{\mu\nu}$ is the difference between 
the full metric of the perturbed spacetime,
and the background. 
The DeWh approach ephasises that the motion is a geodesic of the metric $g_{\mu\nu} + h_{\mu\nu}^{\rm R} $ where 
$h_{\mu\nu}^{\rm R} $ is the radiative part of the perturbation $h_{\mu\nu}$, and it implies two notable features: the regularity of the radiative field and the
avoidance of any non-causal behaviour. The quantitative definition of $h_{\mu\nu}^{\rm R} $ necessitates a brief reminder \cite{popove11}.   
The trace-reversed potential is given by  
\beq
\gamma_{\mu\nu} = h_{\mu\nu} - \frac{1}{2} \bigl(
g^{\rho\sigma} h_{\rho\sigma} \bigr) g_{\mu\nu}~~, 
\label{trpot}
\eeq 
and satisfies the wave equation with the harmonic gauge condition 
\beq 
\Box \gamma^{\mu\nu} + 2  R_{\rho\ \sigma}^{\ \mu\ \nu} 
\gamma^{\rho\sigma} = -16\pi T^{\mu\nu} 
~~~~~~~~~~~~~~~~~~~~~~~~~~~~~~~~~~~~~~\gamma^{\mu\nu}_{\ \ \ ;\nu} = 0~,
\label{box}
\eeq
where $\Box = g^{\alpha\beta} \nabla_\alpha \nabla_\beta$, the wave 
operator for the background spacetime, and $T^{\mu\nu}$ is the  
energy-momentum tensor of the point mass given by a Dirac
distribution supported on the particle trajectory $p$. The
retarded solution is  
\begin{equation} 
\gamma^{\mu\nu} = 4 m \int_{p} 
G^{{\rm ret}\ \mu\nu}_{\ \rho\sigma} u^\rho u^\sigma\, d\tau~~, 
\label{1.9.3}
\end{equation}
where $G^{{\rm ret}\ \mu\nu}_{\ \rho\sigma}$ is the retarded Green
function associated with Eq. (\ref{box}). Inversion of Eq. (\ref{trpot}) provides 
the perturbation $h_{\alpha\beta}$. 
Herein, $\gamma^{\rm adv}_{\mu\nu}$ and $\gamma^{\rm ret}_{\mu\nu}$ satisfy Eq. (\ref{box}), while 
$\gamma^{\rm R}_{\mu\nu}$ is
a free gravitational field that satisfies the homogeneous wave
equation associated to Eq. (\ref{box}). Finally, 

\beq 
h^{\rm R}_{\mu\nu;\lambda} = -4 m \Bigl( u_{(\mu}
R_{\nu)\rho\lambda\sigma} + R_{\mu\rho\nu\sigma} u_\lambda \Bigr) u^\rho
u^\sigma + \nabla_\lambda h^{\rm tail}_{\mu\nu}~~, 
\label{hrad}
\eeq
where the tail term is given by 
\beq
h^{\rm tail}_{\mu\nu} = 4 m \int_{-\infty}^{\tau^-}
 \biggl( G^{\rm ret}_{\mu\nu\mu'\nu'}
- \frac{1}{2} g_{\mu\nu} G^{{\rm ret}\ \ \rho}_{\ \rho\mu'\nu'}
\biggr) \bigl( z(\tau), z(\tau')\bigr) u^{\mu'} u^{\nu'}\, d\tau'~~. 
\label{htail}
\eeq     

Equation \ref{htail} displays the current position $z(\tau)$ of the particle,
at which the unprimed tensors are evaluated, as well as
all prior positions $z(\tau')$, at which primed tensors 
are evaluated. The integral is cut short at $\tau^-$ to avoid the singular behaviour of the retarded
Green's function at coincidence. Finally, $u^{\mu'} = dz^{\mu}/d\tau'$. 

\section{The equations of motion in the two metrics and their difference}

A particle, of $z^\alpha$ coordinates, is moving in the background metric $g_{\mu\nu}$. Its geodesic is given by

\beq
\frac{D^2z^\alpha}{d\tau^2}= 
\frac{d u^\alpha}{d\tau } + \Gamma^\alpha_{\mu\nu} u^\mu u^\nu 
=0~~,                                                
\label{bg-ta}
\eeq
where $\tau$, $\Gamma^\alpha_{\mu\nu}$, $u^\alpha=dz^\alpha/d\tau$ are the proper time, Christoffel symbol and four-velocity in the background metric $g_{\mu\nu}$, respectively. Let us now consider the same particle moving in a perturbed metric. 
We define $\hat{z}^\alpha$ as the coordinates of the particle in the full metric ${\hat g}_{\mu\nu} = g_{\mu\nu} + h_{\mu\nu}^{\rm R} $. 
Obviously, the gauge freedom allows to choose a comoving coordinate frame where no acceleration occurs. 
The geodesic is then given by
\beq
\frac{D^2\hat{z}^\alpha}{d\lambda^2}= 
\frac{d \hat{u}^\alpha}{d\lambda } + {\hat \Gamma}^\alpha_{\mu\nu} {\hat u}^\mu {\hat u}^\nu = 
0~~,
\label{fg-ta}
\eeq
where $\lambda$, ${\hat \Gamma}^\alpha_{\mu\nu}$, $\hat{u}^\alpha =d\hat{z}^\alpha/d\lambda$ are the proper time, Christoffel symbol and four-velocity in the full metric, respectively.

We wish to compute the difference in motion between the two geodesics. To this end, we first map the geodesic in the full spacetime, Eq. (\ref{fg-ta}), onto the background spacetime. The mapping will produce an equation which is not any longer geodesic. 
Furthermore, the projection is not uniquely defined, being dependent on the gauge in which $h_{\mu\nu}^{\rm R}$ is represented.  For a discussion on mapping see \cite{sabade08}.

\subsection{Mapping on the background spacetime}

Conversely to the labelling of the coordinates of the particle in the geodesics, Eqs. (\ref{bg-ta},\ref{fg-ta}), the two spacetimes - background and full - are mapped by two equal coordinate systems in the limit  $h\to 0$. Thus

\beq 
d\tau^2= - g_{\mu\nu}dx^\mu dx^\nu~~,
\label{bm} 
\eeq

and 

\beq 
d\lambda^2= - \hat{g}_{\mu\nu}dx^\mu dx^\nu = - 
\left(g_{\mu\nu} + h_{\mu\nu}^{\rm R} \right)
dx^\mu dx^\nu~~.
\label{fm} 
\eeq

Using the relations 

\beq
\frac{d}{d\lambda} = \frac{d\tau}{d\lambda} \frac{d}{d\tau}
~~~~~~~~~~~~~~~~~~~~~~
\frac{d^2}{d\lambda^2} = \left(\frac{d\tau}{d\lambda}\right)^2 \frac{d^2}{d\tau^2}+ \frac{d^2\tau}{d\lambda^2}\frac{d}{d\tau}~~,
\label{relationstaulambda}
\eeq 
Eq. (\ref{fg-ta}) becomes

\beq
\frac{d^2 \hat{z}^\alpha}{d\tau ^2} + 
\frac{d^2\tau}{d\lambda^2}\frac{d\hat{z}^\nu}{d\tau}  +
{\hat \Gamma}^\alpha_{\mu\nu} \frac{d\hat{z}^\mu}{d\tau}\frac{d\hat{z}^\nu}{d\tau}=0~~; 
\label{fgintau}
\eeq
dividing by $(d\tau/d\lambda)$, we get  

\beq
\frac{d^2 \hat{z}^\alpha}{d\tau ^2} + \Gamma^\alpha_{\mu\nu} \frac{d\hat{z}^\mu}{d\tau}\frac{d\hat{z}^\nu}{d\tau} = 
- \Delta \Gamma^\alpha_{\mu\nu} \frac{d\hat{z}^\mu}{d\tau}\frac{d\hat{z}^\nu}{d\tau} - {\cal K} \frac{d\hat{z}^\alpha}{d\tau}~,
\label{fgintau}
\eeq
where ${\cal K} = (d^2\tau/d\lambda^2) (d\lambda/d\tau)^2$, and $\Delta \Gamma^\alpha_{\mu\nu}={\hat \Gamma}^\alpha_{\mu\nu}  - \Gamma^\alpha_{\mu\nu}$. From the latter, we get

\beq
\begin{aligned}
\Delta\Gamma^\alpha_{\mu\nu}\frac{d\hat{z}^\mu}{d\tau}\frac{d\hat{z}^\nu}{d\tau} =  &
\left [\frac{1}{2}\hat{g}^{\alpha\sigma}
\left(\hat{g}_{\mu\sigma ,\nu} + \hat{g}_{\nu\sigma ,\mu} - \hat{g}_{\mu\nu ,\sigma}\right)
- \Gamma^\alpha_{\mu\nu} \right ] \frac{d\hat{z}^\mu}{d\tau}\frac{d\hat{z}^\nu}{d\tau} =  \\
& \left[ 
\frac{1}{2}g^{\alpha\beta}g^{\sigma\rho}h_{\beta\rho}^{\rm R} 
\left(g_{\mu\sigma ,\nu} + g_{\nu\sigma ,\mu} - g_{\mu\nu ,\sigma}\right)
+
\frac{1}{2}g^{\alpha\sigma}
\left(2h_{\mu\sigma ,\nu}^{\rm R}  - h_{\mu\nu ,\sigma}^{\rm R} \right)
\right] \frac{d\hat{z}^\mu}{d\tau}\frac{d\hat{z}^\nu}{d\tau}= \\
&
\left[
\frac{1}{2}g^{\alpha\sigma}
\left(2h_{\mu\sigma ,\nu}^{\rm R}  - h_{\mu\nu ,\sigma}^{\rm R}  - 2 \Gamma^\rho_{\mu\nu}h_{\sigma\rho}^{\rm R} \right)
\right]\frac{d\hat{z}^\mu}{d\tau}\frac{d\hat{z}^\nu}{d\tau}~~. 
\end{aligned}
\label{fgammahatutau}
\eeq

Using the definition of the covariant derivative 

\beq
2 h^{\rm R} _{\mu\sigma;\nu} - h^{\rm R} _{\mu\nu;\sigma} = 
2 h^{\rm R} _{\mu\sigma,\nu} - 2\Gamma^\rho_{\nu\sigma}h^{\rm R} _{\rho\mu} - 2\Gamma^\rho_{\nu\mu}h^{\rm R} _{\rho\sigma}
-   
h^{\rm R} _{\nu\mu,\sigma} + \Gamma^\rho_{\mu\sigma}h^{\rm R} _{\nu\rho} + \Gamma^\rho_{\sigma\nu}h^{\rm R} _{\mu\rho} = 
2 h^{\rm R} _{\mu\sigma,\nu} - h^{\rm R} _{\mu\nu,\sigma} - 2 \Gamma^\rho_{\mu\nu}h^{\rm R} _{\sigma\rho}~, 
\label{defcd}
\eeq
Eq. (\ref{fgammahatutau},\ref{fgintau}) become respectively 

\beq
\Delta\Gamma^\alpha_{\mu\nu} \frac{d\hat{z}^\mu}{d\tau}\frac{d\hat{z}^\nu}{d\tau} = \left[
\frac{1}{2} g^{\alpha\sigma}
\left(2 h_{\mu\sigma ;\nu}^{\rm R}  - h_{\mu\nu ;\sigma}^{\rm R}  \right)
\right]\frac{d\hat{z}^\mu}{d\tau}\frac{d\hat{z}^\nu}{d\tau}~, 
\label{chris-f}
\eeq

\beq
\frac{D^2\hat{z}^\alpha}{d\tau^2}= -
\frac{1}{2} g^{\alpha\sigma}
\left(2 h_{\mu\sigma ;\nu}^{\rm R}  - h_{\mu\nu ;\sigma}^{\rm R}  \right)
\frac{d\hat{z}^\mu}{d\tau}\frac{d\hat{z}^\nu}{d\tau}
- {\cal K} \frac{d\hat{z}^\alpha}{d\tau}~~.
\label{fgintauexpl}
\eeq
 
The velocity and acceleration vector are orthogonal to each other. We now project Eq. (\ref{fgintauexpl}) orthogonally to 
${d\hat{z}^\alpha}/{d\tau}$ and obtain

\beq
\begin{aligned}
\frac{D^2\hat{z}^\alpha}{d\tau^2} = 
\frac{d^2 \hat{z}^\alpha}{d\tau ^2} + \Gamma^\alpha_{\mu\nu} \frac{d\hat{z}^\mu}{d\tau}\frac{d\hat{z}^\nu}{d\tau} =  
&
- \frac{1}{2} g^{\beta\sigma}
\left(2 h_{\mu\sigma ;\nu}^{\rm R}  - h_{\mu\nu ;\sigma}^{\rm R}  \right) 
\left(\delta^\alpha_\beta + \frac{d\hat{z}^\alpha}{d\tau}\frac{d\hat{z}_\beta}{d\tau}\right) 
\frac{d\hat{z}^\mu}{d\tau}\frac{d\hat{z}^\nu}{d\tau} =\\
& 
- \frac{1}{2} g^{\beta\sigma}
\left(\delta^\alpha_\beta + \frac{d\hat{z}^\alpha}{d\tau}\frac{d\hat{z}_\beta}{d\tau} \right) 
\left(2 h_{\mu\sigma ;\nu}^{\rm R}  - h_{\mu\nu ;\sigma}^{\rm R}  \right) 
\frac{d\hat{z}^\mu}{d\tau}\frac{d\hat{z}^\nu}{d\tau} = \\
& 
- \frac{1}{2} 
\left(g^{\alpha\sigma} + \frac{d\hat{z}^\alpha}{d\tau}\frac{d\hat{z}^\sigma}{d\tau} \right) 
\left(2 h_{\mu\sigma ;\nu}^{\rm R}  - h_{\mu\nu ;\sigma}^{\rm R}  \right) 
\frac{d\hat{z}^\mu}{d\tau}\frac{d\hat{z}^\nu}{d\tau}~~. 
\label{fgintauexplfin}
\end{aligned}
\eeq

Equation (\ref{fgintauexplfin}) couples the particle full coordinates with the background proper time.  

\subsection{Difference between the motions of a test-particle and of a particle affected by its own mass and radiation}

The framework entails that the motion and the self-force are computed up to first order in $h$.
Assuming $\hat{z}^\alpha = z^\alpha + \Delta z^\alpha$, the left-hand side of Eq. (\ref{fgintauexplfin}) can be rewritten as 

\beq
\frac{d^2 (z^\alpha + \Delta z^\alpha)}{d\tau ^2} + 
\Gamma^\alpha_{\mu\nu} (z^\alpha + \Delta z^\alpha) 
\frac{d(z^\mu + \Delta z^\mu)}{d\tau} \frac{d(z^\nu + \Delta z^\nu)}{d\tau}~~. 
\eeq 

The Christoffel symbol may be expanded as 

\beq
\Gamma^\alpha_{\mu\nu} (z^\alpha + \Delta z^\alpha) = \Gamma^\alpha_{\mu\nu} (z^\alpha) + \Gamma^\alpha_{\mu\nu,\rho} \Delta z^\rho + O (h^2)~~. 
\eeq

We then obtain up to first order  

\[
\frac{d^2 z^\alpha}{d\tau ^2} + \Gamma^\alpha_{\mu\nu} \frac{dz^\mu}{d\tau}\frac{dz^\nu}{d\tau} + 
\frac{d^2 \Delta z^\alpha}{d\tau ^2} + 2 
\Gamma^\alpha_{\mu\nu} \frac{d \Delta z^\mu}{d\tau}\frac{d z^\nu}{d\tau} 
+ \Gamma^\alpha_{\mu\nu,\rho} \frac{dz^\mu}{d\tau}\frac{dz^\nu}{d\tau} \Delta z^\rho
= 
\]
\beq
- \frac{1}{2}
\left( g^{\alpha\beta} + \frac{dz^\alpha}{d\tau}\frac{dz^\beta}{d\tau} \right) 
(2h_{\mu\beta ;\nu}^{\rm R} - h_{\mu\nu ;\beta}^{\rm R} )\frac{dz^\mu}{d\tau}\frac{dz^\nu}{d\tau} 
~, 
\eeq
where the first two terms on the left-hand side correspond to the geodesic of Eq. (\ref{bg-ta}). Thus 
(${dz^\mu}/{d\tau} = u^\mu$) 

\beq
\frac{d^2 \Delta z^\alpha}{d\tau ^2} = 
- \Gamma^\alpha_{\mu\nu,\rho} u^\mu \Delta z^\rho u^\nu 
- 2 \Gamma^\alpha_{\mu\nu} \frac{d \Delta z^\mu}{d\tau} u^\nu 
- \frac{1}{2}
(g^{\alpha\beta} + u^\alpha u^\beta) 
(2h_{\mu\beta ;\nu}^{\rm R} - h_{\mu\nu ;\beta}^{\rm R} )u^\mu u^\nu 
~~, 
\label{d2Deltazdtau2}
\eeq

The next step consists in transforming Eq. (\ref{d2Deltazdtau2}) into a covariant form. 
Introducing the covariant derivatives, it holds that

\[
\frac{D^2 \Delta z^\alpha}{d\tau^2}= 
\frac{D}{d\tau}\left(\frac{D \Delta z^\alpha}{d\tau}\right) = 
\frac{d}{d\tau}
\left( \frac{d \Delta z^\alpha}{d\tau} + \Gamma^\alpha_{\mu\nu}u^\mu \Delta z^\nu\right) 
+  
\Gamma^\alpha_{\sigma\rho}
\left(\frac{d\Delta z^\sigma}{d\tau}+ \Gamma^\sigma_{\mu\nu}u^\mu \Delta z^\nu\right)u^\rho =
\]
\beq
\frac{d^2 \Delta z^\alpha}{d\tau^2}     
+ \Gamma^\alpha_{\mu\nu,\rho}u^\mu \Delta z^\nu u^\rho
+ 2 \Gamma^\alpha_{\mu\nu} \frac{d \Delta z^\mu}{d\tau}u^\nu 
+ \Gamma^\alpha_{\sigma\rho}~\Gamma^\sigma_{\mu\nu}u^\rho \Delta z^\mu u^\nu + O (h^2)~~; 
\eeq     
we then make use of Eq. (\ref{d2Deltazdtau2}) and arrive to the covariant correction at the first order of the background geodesic

\beq
\frac{D^2 \Delta z^\alpha}{d\tau^2}=
\left(
~\Gamma^\alpha_{\mu\beta,\nu} - ~\Gamma^\alpha_{\mu\nu,\beta}
+ ~\Gamma^\alpha_{\sigma\nu}~\Gamma^\sigma_{\mu\beta} 
- ~\Gamma^\alpha_{\sigma\beta}~\Gamma^\sigma_{\mu\nu} 
\right)
u^\mu \Delta z^\beta u^\nu  -
\frac{1}{2}(g^{\alpha\beta} + u^\alpha u^\beta) (2h_{\mu\beta ;\nu}^{\rm R}  - h_{\mu\nu ;\beta}^{\rm R} ) u^\mu u^\nu~~. 
\label{gweq-1}
\eeq

Having recognised the Riemann tensor, we have

\beq
\frac{D^2 \Delta z^\alpha}{d\tau^2}=
- {R_{\mu\beta\nu}}^\alpha u^\mu \Delta z^\beta u^\nu 
- \frac{1}{2}
(g^{\alpha\beta} + u^\alpha u^\beta) 
(2h_{\mu\beta ;\nu}^{\rm R} - h_{\mu\nu ;\beta}^{\rm R} ) u^\mu u^\nu~~. 
\label{almostgweq}
\eeq

The Riemann tensors in Eq. (\ref{hrad}) disappear when $h^{\rm R}$ is replaced in Eq. (\ref{almostgweq}) \cite{popove11}; we finally get
\beq
\frac{D^2 \Delta z^\alpha}{d\tau^2}=
\underbrace{- {R_{\mu\beta\nu}}^\alpha u^\mu \Delta z^\beta u^\nu}_{Background~metric~geodesic~deviation} 
\underbrace{- \frac{1}{2}
(g^{\alpha\beta} + u^\alpha u^\beta) 
(2h_{\mu\beta ;\nu}^{\rm tail} - h_{\mu\nu ;\beta}^{\rm tail} ) u^\mu u^\nu}_{Self-acceleration}~~. 
\label{gweq}
\eeq

Stemmed from geodesic principles, a geodesic deviation equation is thus obtained by subtracting the background from the perturbed motion, Eq. (\ref{gweq}). This equation has appeared first in \cite{grwa08,grwa11}, with a different derivation.  The first right-hand side term depends on the background metric, while the second right-hand term depends upon the perturbations and it is the non-trivial 
self-acceleration term. The latter, multiplied by $m$, provides the known MiSaTaQuWa equation \cite{misata97, quwa97}.

The interpretation of Eq.(\ref{gweq}) leads to consider the self-acceleration term causing a displacement in the trajectory represented by the geodesic deviation in the background metric. 

\section{Discussion and conclusions}

The evolution of an orbit is object of growing interest in the frame of an SLI mission, since the self-force affects the waveforms of EMRIs through dephasing, and thus it is to be taken into account for a successful detection.  Equation (\ref{gweq}) is the exact expression of a perturbative approach, but for the evolution of an orbit, a self-consistent approach \cite{grwa08,grwa11} is preferable. 
This approach prescribes that the self-acceleration term be continuously applied all along the background trajectory and thereby correcting it, neglecting the geodesic deviation term. At each successive instant, the geodesic is corrected by the self-acceleration, and a new obsculating geodesic is determined.   
(Quasi-)circular orbits and inspirals, have been evolved self-consistently for the scalar and gravitational cases \cite{divewade12,labu12,waakbagasa12}, as the radial gravitational infall \cite{spri14}.   
 
Herein, we have proposed a simple derivation of the first order perturbative correction to the geodesic of the background in a covariant form. We have found this approach instructive when dwelling on the significance of the motion of a particle in a curved background. 

\section*{Appendix: an alternative way} 

Herein, we trace a different path also leading to Eq. (\ref{gweq}). One of the differences is the avoidance of the projection adopted to get to Eq. (\ref{fgintauexplfin}). 
First, we write the relation between the proper times. 
From Eqs. (\ref{bm},\ref{fm}), we have   

\beq 
d\lambda^2 = d\tau^2 - h_{\mu\nu}^{\rm R}  dx^\mu dx^\nu~,
\label{fm-bm-ta} 
\eeq

\beq 
\frac{d\lambda}{d\tau} = 1 - \frac{1}{2} h_{\mu\nu}^{\rm R}  u^\mu u^\nu
~~~~~~~~~~
\frac{d\tau}{d\lambda} = 1 + \frac{1}{2} h_{\mu\nu}^{\rm R}  u^\mu u^\nu~~,
\label{fmbmratio-ta} 
\eeq

\beq
\frac{d^2\tau}{d\lambda^2} = \frac{d\tau}{d\lambda} \frac{d}{d\tau}\frac{d\tau}{d\lambda} = 
\frac{1}{2} h_{\mu\nu;\sigma}^{\rm R}  u^\mu u^\nu u^\sigma~~,   
\label{d2taudlambda2-ta}
\eeq
having neglected $h_{\mu\nu}^{\rm R}  a^\mu u^\nu$, being the acceleration of order $h$. Second, we determine the geodesic in the full metric in terms of the background affine parameter. We consider 

\[
{\hat \Gamma}^\alpha_{\mu\nu}\hat{u}^\mu \hat{u}^\nu = 
\frac{1}{2}\hat{g}^{\alpha\sigma}
\left(\hat{g}_{\mu\sigma ,\nu} + \hat{g}_{\nu\sigma ,\mu} - \hat{g}_{\mu\nu ,\sigma}\right)\hat{u}^\mu \hat{u}^\nu
=
\]
\[
\left[
\Gamma^\alpha_{\mu\nu} 
- \frac{1}{2}g^{\alpha\beta}g^{\sigma\rho}h_{\beta\rho}^{\rm R} 
\left(g_{\mu\sigma ,\nu} + g_{\nu\sigma ,\mu} - g_{\mu\nu ,\sigma}\right)
+
\frac{1}{2}g^{\alpha\sigma}
\left(2h_{\mu\sigma ,\nu}^{\rm R}  - h_{\mu\nu ,\sigma}^{\rm R} \right)
\right] \hat{u}^\mu \hat{u}^\nu =
\]
\beq
\left[
\Gamma^\alpha_{\mu\nu} 
+ \frac{1}{2}g^{\alpha\sigma}
\left(2h_{\mu\sigma ,\nu}^{\rm R}  - h_{\mu\nu ,\sigma}^{\rm R}  - 2 \Gamma^\rho_{\mu\nu}h_{\sigma\rho}^{\rm R} \right)
\right]\hat{u}^\mu \hat{u}^\nu~~. 
\label{fgammahatu}
\eeq

Equation (\ref{fgammahatutau}) differs from Eq. (\ref{fgammahatu}), the former referring to ${d\hat{z}^\mu}/{d\tau}$, the latter
to ${d\hat{z}^\mu}/{d\lambda}$. Using Eq. (\ref{defcd}), we get   

\beq
{\hat \Gamma}^\alpha_{\mu\nu} \hat{u}^\mu \hat{u}^\nu = \left[
\Gamma^\alpha_{\mu\nu} 
+ \frac{1}{2}g^{\alpha\sigma}
\left(2h_{\mu\sigma ;\nu}^{\rm R}  - h_{\mu\nu ;\sigma}^{\rm R}  \right)
\right]\hat{u}^\mu \hat{u}^\nu~~.
\label{chris-f-ta}
\eeq

Now, using Eq. (\ref{relationstaulambda}), we rewrite Eq. (\ref{fg-ta}) in terms of the background spacetime parameters

\beq
\frac{d^2 \hat{z}^\alpha}{d\lambda ^2} + {\hat \Gamma}^\alpha_{\mu\nu} 
\frac{d\hat{z}^\mu}{d\lambda}\frac{d\hat{z}^\nu}{d\lambda} =
\left(\frac{d\tau}{d\lambda}\right)^2 \frac{d^2 \hat{z}^\alpha}{d\tau^2} 
+ 
\frac{d^2\tau}{d\lambda^2}\frac{d\hat{z}^\alpha}{d\tau}
+
{\hat \Gamma}^\alpha_{\mu\nu} \left(\frac{d\tau}{d\lambda}\right)^2 
\frac{d\hat{z}^\mu}{d\tau}\frac{d\hat{z}^\nu}{d\tau} 
~~.
\label{fg(b)-ta}
\eeq

We divide Eq. (\ref{fg(b)-ta}) by $(d\tau/d\lambda)^2$; using Eqs. (\ref{fmbmratio-ta},\ref{d2taudlambda2-ta}), and truncating at first order, we have 
that the middle term on the right-hand side transforms into 

\beq
\frac{d^2\tau}{d\lambda^2}\frac{d\hat{z}^\alpha}{d\tau} \left(\frac{d\lambda}{d\tau}\right)^2 = \frac{1}{2} h_{\mu\nu;\sigma}^{\rm R}  u^\mu u^\nu u^\sigma \frac{d\hat{z}^\alpha}{d\tau}~~.
\eeq

Finally, we have the geodesic in the full metric in terms of the background affine parameter

\beq
\frac{d^2 \hat{z}^\alpha}{d\tau^2} 
+ 
\frac{1}{2} h_{\mu\nu;\sigma}^{\rm R}  u^\mu u^\nu u^\sigma \frac{d\hat{z}^\alpha}{d\tau}
+
{\hat \Gamma}^\alpha_{\mu\nu} \frac{d\hat{z}^\mu}{d\tau}\frac{d\hat{z}^\nu}{d\tau}=0~~.
\label{gfmbap-ta}
\eeq

Third, in Eq. (\ref{gfmbap-ta}), we explicit the coordinate where the Christoffel symbol refers to, and get

\[
\frac{d^2 \hat{z}^\alpha}{d\tau^2} 
+ 
\frac{1}{2} h_{\mu\nu;\sigma}^{\rm R}  u^\mu u^\nu u^\sigma \frac{d\hat{z}^\alpha}{d\tau}
+
{\hat \Gamma}^\alpha_{\mu\nu} (\hat{z}^\alpha) 
\frac{d\hat{z}^\mu}{d\tau}\frac{d\hat{z}^\nu}{d\tau}
= 
\]
\beq
\frac{d^2 z^\alpha}{d\tau^2} 
+
\frac{d^2 \Delta z^\alpha}{d\tau^2} 
+
\frac{1}{2} h_{\mu\nu;\sigma}^{\rm R}  u^\mu u^\nu u^\sigma u^\alpha
+
{\hat \Gamma}^\alpha_{\mu\nu}(z^\alpha + \Delta z^\alpha)  u^\mu u^\nu
+ 
2 \Gamma^\alpha_{\mu\nu}(z^\alpha)  u^\mu \frac{d\Delta z^\nu}{d\tau}~~.
\eeq 

Now, we approximate the Christoffel symbol in the full metric (${\hat \Gamma}^\alpha_{\mu\nu,\rho}\Delta z^\rho$ is of order $h^2$)
\beq
{\hat \Gamma}^\alpha_{\mu\nu}(z^\alpha + \Delta z^\alpha) = ~{\hat \Gamma}^\alpha_{\mu\nu}(z^\alpha) + 
\Gamma^\alpha_{\mu\nu,\rho}\Delta z^\rho + O(h^2)~,
\eeq 
and get, using Eqs. (\ref{bg-ta},\ref{chris-f-ta})

\beq
\frac{d^2 \Delta z^\alpha}{d\tau^2} 
+ 
g^{\alpha\beta}
\left(h_{\mu\beta ;\nu}^{\rm R}  - \frac{1}{2}h_{\mu\nu ;\beta}^{\rm R}  \right) u^\mu u^\nu
+
\Gamma^\alpha_{\mu\nu,\rho}u^\mu u^\nu \Delta z^\rho +
2 \Gamma^\alpha_{\mu\nu}\frac{d\Delta z^\mu}{d\tau}u^\nu 
+
\frac{1}{2} h_{\mu\nu;\sigma}^{\rm R}  u^\mu u^\nu u^\sigma u^\alpha = 0~~.
\eeq 

We then step into Eq. (\ref{d2Deltazdtau2}), and from then on, we recover the steps already described.

\bibliography{references_spallicci_140501}

\end{document}